\documentclass[toc]{PoS}
\title{Dilepton measurements with HADES}
\ShortTitle{Dilepton measurements with HADES}
\author{\speaker{Christian Müntz}%
         \thanks{for the HADES Collaboration}\\
        J. W. Goethe-Universität Frankfurt\\
        E-mail: \email{c.muentz@gsi.de}}

\abstract{HADES is the only operating dielectron spectrometer in the
energy regime 1-2~AGeV. The physics program aims at a systematic
investigation of dielectron production in heavy ion
as well as elementary and pion-induced reactions.\\
This contribution highlights recent results on electron pair
production in $^{12}$C+$^{12}$C collisions at an incident energy of
2~AGeV with HADES. The measured pair production probabilities span
over five orders of magnitude. Dalitz decays of $\pi^0$ and $\eta$
account for all the yield up to 0.15 GeV/c$^2$, but for only about
50\% above this invariant mass. The excess yield compared to the
hadronic cocktail between the $\pi^0$-Dalitz and the $\rho/\omega$
invariant-mass region is in agreement with the former DLS result if
one assumes that it scales with beam energy like pion production.
Preliminary results from $^{12}$C+$^{12}$C collisions at an incident
energy of 1~AGeV support this scenario.}

\FullConference{Critical Point and Onset of Deconfinement
- 4th International Workshop\\
July 9 - 13, 2007\\
Darmstadt, Germany}

\begin{document}

\section{The mission of HADES}
The properties of hot and dense hadronic matter represent a key
problem in heavy- ion physics, with far-reaching implications for
other fields such as the physics of compact stars. They are governed
by non-perturbative QCD. Models predict, however, that hadron
properties, such as mass and lifetime, depend on the temperature and
density of the surrounding nuclear medium. While some hadronic
many-body calculations give a broadening of the meson in-medium
spectral function, other approaches predict dropping meson masses
related to the restoration of chiral symmetry~\cite{general}. The
experimental access is difficult as well, since the life time of the
dense phase is extremely short, and most of the probes suffer from
final state interactions, substantially masking the primary
information. Dileptons ($\mu^+ \mu^-$ or $e^+e^-$) from decays of
short-lived resonances produced inside the hadronic matter created
in the course of (ultra)-relativistic heavy-ion collisions, however,
are considered to represent the most direct probe, and dedicated
experiments and analyses, e.g. CERES and NA60 at CERN SPS, and
PHENIX at RHIC~\cite{CPOD}, are presently promoted. In the 1-2~AGeV
incident energy regime the small branching ratios into the
dielectron channel adds a particular experimental difficulty. The
DLS collaboration found unexpectedly large electron-pair yields in
C+C and Ca+Ca collisions~\cite{DLS} which cannot be described
satisfactorily within the various scenarios proposed for possible
changes of the in-medium spectral functions~\cite{general,mod1}.
Indeed, the pair yields in the invariant-mass range between 0.15 and
0.6 GeV/c$^2$, i.e.~just below the $\rho$ meson pole mass, still
remain to be explained~\cite{mod2}. Resolving this unsatisfactory
situation by means of a high-accuracy and systematic investigation
of dielectron production in heavy ion as well as elementary
reactions was the primary motivation to set up the
$2^{nd}$-generation experiment HADES

\section{The HADES setup}
\begin{figure}[h]
  \centering
       \centering
     \begin{minipage}[c]{0.40\linewidth}
 \centering
        \caption[]{
Cross section of the HADES setup, with tracks of charged particles,
including an electron which triggers Cherenkov light in the RICH and
an electromagnetic shower in the Pre-Shower detector in the upper
sector. The beam is coming from the left, the RICH diameter amounts
to 1.6~m, see text for details.
  }\label{FigHades}
  \end{minipage}
     \hspace{0.02\linewidth}
     \begin{minipage}[c]{0.38\linewidth}
       \includegraphics[width=\linewidth]{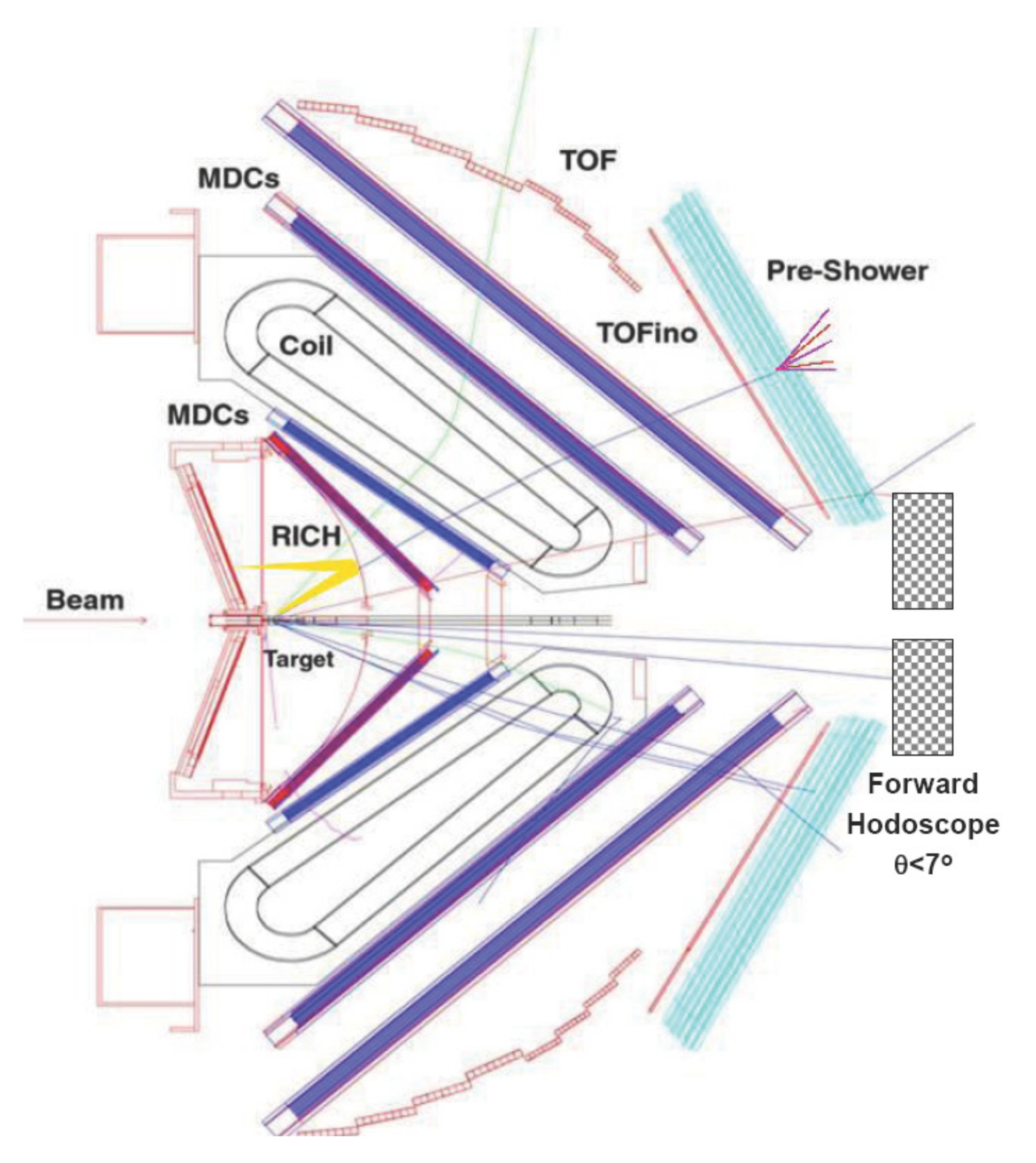}
    \end{minipage}
\end{figure}

The High-Acceptance DiElectron Spectrometer HADES at GSI, Darmstadt,
is presently the only dilepton spectrometer operating in the energy
regime of 1-2~AGeV, succeeding the DLS spectrometer at
Bevalac~\cite{DLS}. The HADES spectrometer, described in detail
in~\cite{HADES}, consists of a 6-coil toroidal magnet centered on
the beam axis and six identical detection sections located between
the coils, covering polar angles between $18^{\circ}$ and
$85^{\circ}$, see figure~\ref{FigHades} where a cross section of the
detector setup is depicted. In the measurement $^{12}$C+$^{12}$C at
2~AGeV, each sector was composed of a gaseous Ring-Imaging Cherenkov
(RICH) detector, two planes of Mini-Drift Chambers (MDC-I and
MDC-II) for track reconstruction between the RICH and the magnetic
field, and a Time-Of-Flight wall (TOF/TOFino) supplemented at
forward polar angles with Pre-SHOWER chambers. The interaction time
was obtained from a fast diamond start detector located upstream of
the target. In the following experiments ($^{12}$C+$^{12}$C at
1~AGeV,  $^{40}$Ar+KCl at 1.75~AGeV and elementary reactions) the
tracking system was bit by bit completed with two planes of drift
chambers between the magnetic field and the time-of-flight system,
aiming at maximum momentum resolution in all six sectors. Hence,
HADES provides a large and smooth dielectron acceptance of about
\mbox{30\%}. Recently, the Forward Hodoscope was added under polar
angels smaller than $7^{\circ}$, which serves for tagging pn
reactions in d-induced reactions and which will provide information
on the reaction plane orientation in future heavy ion runs. Besides
solid state targets, a liquid hydrogen target is available for the
physics program regarding elementary reactions.
For C+C (2 AGeV) the data readout was started by a first-level
trigger (LVL1) decision, requiring a charged-particle multiplicity
$MUL\leq 4$ in the TOF/TOFINO detectors, accepting 60~\% of the
total cross section. It was followed by a second-level trigger
(LVL2) requesting at least one electron track. With this trigger
condition, a 10-fold pair enrichment at a pair efficiency of $\leq
92$~\% was achieved. Furthermore, the LVL2 introduced no bias on the
shapes of measured pair distributions, as checked by a direct
comparison to the unbiased LVL1 events. Uncorrelated electrons from
$\pi^0\to\gamma\gamma$ decays followed by photon conversion and/or
from $\pi^0\to e^+e^-\gamma$ Dalitz decays form most of the
combinatorial background (CB). They are rejected efficiently by
applying conditions on the opening angle $\theta_{e+e-}>9^{\circ}$
of the reconstructed track segments, removing $95\%$ of the
conversion pairs while reducing the dielectron signal with
$M_{ee}>0.15$ GeV/c$^2$ by less than $10\%$. Here, all numbers given
are valid for the 2~AGeV $^{12}$C+$^{12}$C run. For further details,
especially regarding the analysis, determining the combinatorial
background as well as systematical errors, see~\cite{2AGeV,HADES}.\\

\section{Experimental results}
\begin{figure}[h]
  \centering
       \centering
     \begin{minipage}[c]{0.35\linewidth}
 \centering
        \caption[]{
(a) Efficiency- and background-corrected $e^+e^-$ invariant-mass
  distribution for $\theta_{e+e-}>9^o$ (symbols) compared to a thermal
  dielectron cocktail of free $\pi^0$, $\eta$ and $\omega$ decays
  (cocktail A, solid line), as well as including $\rho$ and $\Delta$ resonance decays
  (cocktail B, long-dashed line). Only statistical errors are shown.
  (b) Ratio of data and cocktail A (dots), compared to ratios
  of various model calculations and cocktail A as a function of the invariant mass.
  All calculations
  have been filtered and folded with the HADES acceptance and mass
  resolution. Statistical and systematic errors of the measurement
  are shown as vertical and horizontal lines, respectively.
  The overall normalization error of 11\% is depicted by the shaded area.
  }\label{2AGeV}
  \end{minipage}
     \hspace{0.02\linewidth}
     \begin{minipage}[c]{0.6\linewidth}
       \includegraphics[width=\linewidth]{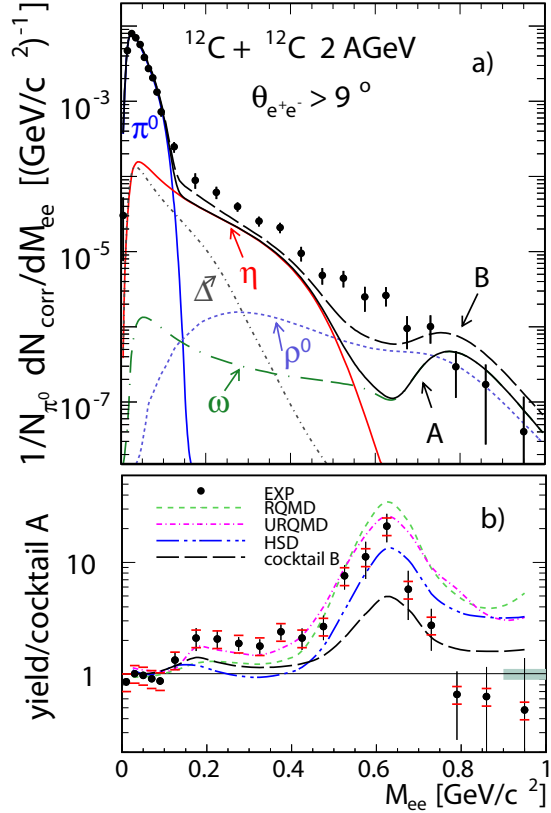}
    \end{minipage}
\end{figure}

Figure~\ref{2AGeV}a shows the $M_{ee}$ distribution of the signal
pairs after efficiency correction and normalization to the average
number of charged pions $N_{\pi}=\frac{1}{2}(N_{\pi^+}+N_{\pi^-})$.
The charged pion yield was measured in the HADES acceptance
\cite{HADES} and extrapolated to the full solid angle. The
extrapolation takes into account measured angular distributions,
found to be in agreement with UrQMD calculations \cite{bleicher}.
The obtained pion multiplicity per number of participating nucleons
$M_{\pi}/A_{part}=0.137\pm0.015$ ($A_{part}=9.0$) agrees with
previous measurements of charged and neutral pions \cite{taps1kaos}
within the quoted error of $11\%$. The error represents an overall
normalization error, dominated by systematic uncertainties in the
acceptance and efficiency corrections of the charged-pion analysis.
The systematical uncertainties for the dielectron yield from the
efficiency correction and the CB subtraction add up quadratically to
a nearly constant error of $18\%$.
A pair cocktail (cocktail A) was calculated from free $\pi^0$,
$\eta$ and $\omega$ meson decays only to represent all contributions
emitted after the chemical freeze-out of the fireball and compared
to the resulting $M_{ee}$ distribution (figure~\ref{2AGeV}a). While
$\pi^0$ and $\eta$ sources are directly constrained by data
\cite{taps1kaos}, the production rate of the $\omega$ meson is taken
from a $m_{\perp}$-scaling ansatz \cite{bratkovskaya}. In the HADES
event generator PLUTO~\cite{PLUTO} meson production was modeled
assuming emission from a thermal source with a temperature
$T=80$~MeV, without a radial expansion velocity. The anisotropic
angular distribution, obtained from the charged-pion analysis, was
used for the $\pi^0$ mesons. While experimental data and the
simulated cocktail A are in good agreement in the $\pi^0$ region,
the cocktail undershoots the data for $M>0.15$ GeV/c$^2$. Decays of
short-lived resonances (as $\rho$, $\Delta(1232)$), excited in the
early phase of the collision, will additionally contribute here.
Assuming that the $\Delta$ yield scales with the $\pi^0$ yield and,
the $\Delta^{0,+}\to Ne^+e^-$ decays were added to the cocktail,
using a calculated decay rate~\cite{ernst}. Furthermore, the $\rho$
meson was treated in analogy to the $\omega$ meson. For the latter
broad resonance ($\Gamma_{\rho}=150$ MeV), $m_{\perp}$ scaling and
available phase space strongly enhances the low-mass tail, resulting
in the skewed shape visible in figure~\ref{2AGeV}a. Comparing the
full thermal cocktail (cocktail B) with the data, the simulated
yield above 0.15 GeV/c$^2$ is now increased and the high-mass region
is populated with dielectrons from $\rho \to e^+e^-$ decays, but the
calculation still falls short of reproducing the data. In
figure~\ref{2AGeV}b the ratio of the data and cocktail A is shown.
In the intermediate mass range of $0.15 - 0.50$ GeV/c$^2$, the
enhancement factor above the dominant $\eta$ contribution amounts to
$F(2.0) = 2.1\pm 0.2(stat) \pm 0.3 (sys) \pm 0.4 (\eta)$. The third
error ($\eta$) indicates the errors resulting from the quoted error
in the measured
$\eta$ multiplicity with TAPS~\cite{taps1kaos}.\\

\begin{figure}[h]
  \centering
       \centering
     \begin{minipage}[c]{0.30\linewidth}
 \centering
        \caption[]{
        Ratio of data (1~AGeV and 2~AGeV incident energy) and the corresponding
        cocktail A of dielectrons originating from long-lived mesons as a function of the invariant mass,
        see text.  The overall normalization error of 11\% is depicted by the shaded area.
  }\label{Ratio1AGeV}
  \end{minipage}
     \hspace{0.02\linewidth}
     \begin{minipage}[c]{0.65\linewidth}
       \includegraphics[width=\linewidth]{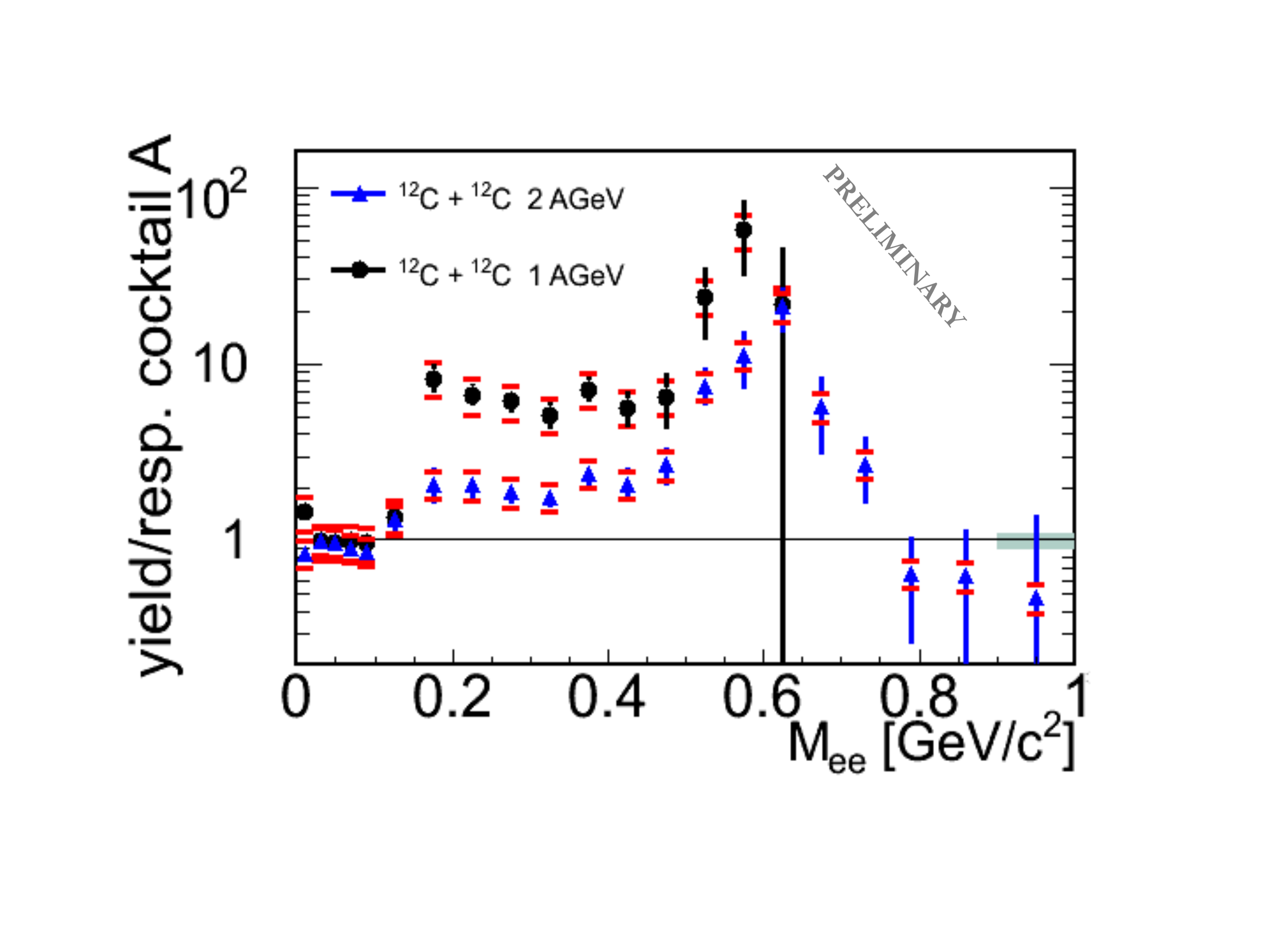}
    \end{minipage}
\end{figure}

The analysis of the 1~AGeV $^{12}$C+$^{12}$C run is presently being
concluded. Hence, only preliminary data can be shown and discussed.
Comparing the acceptance- and efficiency-corrected invariant mass
spectrum measured by HADES to the corresponding DLS
spectrum~\cite{DLS}, by extrapolating to the DLS phase space, an
agreement within the statistical and systematical errors can be
reported~\cite{1AGeV}. In comparison to the 2~AGeV invariant mass
spectrum the excess yield is much more pronounced, see
figure~\ref{Ratio1AGeV}, yielding a preliminary enhancement factor
above the dominant $\eta$ contribution of $F(1.0) = 7.0\pm 0.6(stat)
\pm 1.1 (sys) \pm 2.0 (\eta)$. For generating cocktail A (1~AGeV)
representing the long-lived mesons, a freeze-out temperature of
T=55~MeV was assumed.

Employing the $\eta$-yield measured by TAPS~\cite{taps1kaos} at 1.04
and 2~AGeV the preliminary ratio of the absolute excess yields
$Y_{exc}$ in $^{12}$C+$^{12}$C reactions at both energies can be
deduced: $Y_{exc}^{2AGeV}/Y_{exc}^{1AGeV}$ = $2.8\pm 0.2(stat)$ $\pm
1.0 (sys)$$ \pm 1.0 (\eta)$. It is interesting to note that the
energy scaling of the excess yield is similar to that of the pion
production and does not follow the energy scaling of $\eta$
mesons~\cite{taps1kaos}.

\section{Summary and outlook}
HADES was set up at SIS/GSI Darmstadt to study dielectron production
in heavy ion as well as elementary and pion-induced reactions in the
energy regime of 1-2~AGeV in a very systematic way, with quality
data. This paper highlights final results on $^{12}$C+$^{12}$C
collisions at 2~AGeV incident energy, and their comparison to the
preliminary results from 1~AGeV. In both data sets a significant
dielectron excess yield above the thermal pair cocktail from
decaying long-lived mesons was found. A quantitative analysis of
this excess confirms the results reported by the DLS collaboration
at 1~AGeV. The excess energy dependence is similar to the one of
pions. This observation suggests the importance of baryonic
resonances for the origin of the excess yield. It also demonstrates
the need of a systematic investigation of elementary reactions as
input for models, presently being pursued by the HADES
collaboration. First preliminary results are
available~\cite{elementary}, focusing on the role of bremsstrahlung
by comparing pp with pn (d-induced) reactions at 1.25~GeV and
hunting vector mesons in pp reactions at 3.5~GeV. Regarding the
heavy ion program data from Ar + KCl at 1.75~AGeV are presently
being analyzed, and HADES is facing a major upgrade by installing
multi-gap timing RPC detectors and replacing central DAQ components
to cope with the high multiplicities in Au + Au reactions.

\end{document}